# Solar Eruptions, Forbush Decreases and Geomagnetic Disturbances from Outstanding Active Region 12673

**I. M. Chertok[1], A. V. Belov[1], and A. A. Abunin[1,2]**

[1]Pushkov Institute of Terrestrial Magnetism, Ionosphere and Radio Wave Propagation (IZMIRAN), Troitsk, Moscow 108840, Russia
[2]Kalmyk State University, Elista358000, Russia

Correspondingauthor: Ilya Chertok (ichertok@izmiran.ru)

**Key Points:**

- The total magnetic flux of the EUV arcades and dimmings of two solar eruptive flares from early September 2017 is used for geoeffectiveness diagnostics.
- The estimated scales of the space weather disturbances caused by the flares are close to those of the observed Forbush decreases and geomagnetic storms.
- The roles of a high-speed stream from an adjacent coronal hole and the interaction between two ICMEs near Earth are considered.

## Abstract

Based on our tool for the early diagnostics of solar eruption geoeffectiveness (EDSEG tool; Chertok et al., 2013, 2015, 2017), we have analyzed space weather disturbances that occurred in early September 2017. Two flares, SOL2017-09-04T20:33 (M5.5) and SOL2017-09-06T12:02 (X9.3), accompanied by Earth-directed halo coronal mass ejections (CMEs) were found to be geoeffective. We extracted the associated EUV dimmings and arcades and calculated their total unsigned magnetic flux. This calculation allowed us to estimate the possible scales of the Forbush decreases (FDs) and geomagnetic storms (GMSs) in the range from moderate to strong, and they are close to the observed scales. More precisely, after the first eruption, an FD approximately equal to 2% and almost no GMS occurred because the $Bz$ magnetic field component in front of the corresponding interplanetary CME (ICME) was northern. The stronger second eruption produced somewhat larger composite disturbances (FD ≈ 9.3% and GMS with indexes Dst ≈ −144 nT, $Ap$ ≈ 235) than expected (FD ≈ 4.4%, Dst ≈ −135 nT, $Ap$ ≈ 125) because the second ICME overtook the trailing part of the first ICME near Earth, and the resulting Bz component was more intense and southern. Both ICMEs arrived at Earth earlier than expected because they propagated in the high-speed solar wind emanated from an extended coronal hole adjacent to AR12673 along their entire path. Overall, the presented results provide further evidence that the EDSEG tool can be used for the earliest diagnostics of actual solar eruptions to forecast the scale of the corresponding geospace disturbances.

## Plain Language Summary

Space weather nonrecurrent disturbances such as geomagnetic storms (GMSs) and Forbush decreases (FDs) of galactic cosmic rays are caused by coronal mass ejections (CMEs) that erupt from the Sun and propagate to Earth. We analyze such disturbances in connection with an outstanding outburst of solar flare activity that occurred at the beginning of September 2017 and compare their parameters with estimates from our tool for the early diagnostics of solar eruption geoeffectiveness. The estimates are based on measurements of the magnetic flux of the earliest CME manifestations in the extreme ultraviolet range such as bright coronal post-eruptive arcades and areas with temporarily reduced brightness (so called dimmings). We demonstrate that this tool yields the earliest relatively correct estimates for the scales of GMS and FD despite the complexity of the near-Earth solar wind structures in September 2017.

## 1 Introduction

The current solar cycle 24 has been rather weak, particularly concerning the number of sunspots. Nevertheless, one of the most powerful outbursts of flare activity occurred during the deep decline phase of this cycle in early September 2017 (ftp://ftp.sec.noaa.gov/pub/warehouse/2017/). Within one week, September 4–10, the Sun produced numerous C-class flares, 27 M-class flares, and four major X-class flares, including the two strongest flares in the cycle: SOL2017-09-06 (X9.3) and SOL2017-09-10 (X8.2). The outburst was due to the extremely rapid development and increasing complexity of the magnetic structure of NOAA active region AR12673 during its passage over the western half of the visible disk. Three halo-type coronal mass ejections (CMEs) were recorded over the course of this activity (https://cdaw.gsfc.nasa.gov/CME_list/HALO/halo.html). The effects on the terrestrial space environment were a number of solar energetic particle (SEP) events including one ground-level enhancement (GLE) on September 10, as well as moderate to strong geomagnetic storms and Forbush decreases in galactic cosmic ray flux (see Space Weather Highlights in SWPC PRF 2193 and 2194 at http://legacy-ww.swpc.noaa.gov/weekly/index.html).

Unsurprisingly, this extreme period has attracted much attention from researchers, and many articles have already been published with results from analyses of various aspects of this period. Attie et al. (2018) found precursors between September 1 and 3 seen as a disruption of the moat flow several hours before the onset of strong flux emergence near the main sunspot of AR 12673. A detailed study of the morphological evolution of AR12673 is presented in Yang et al. (2017). Intense development of the AR was shown to occur on September 3 and 4 in the form of the rapid emergence of multiple bipolar patches near and to the east of the main preexisting sunspot. According to Sun and Norton (2017), the emergence rate of this magnetic flux was one of the fastest ever observed. Further complication of the magnetic structure due to flux emergence and cancellation, strong shearing motions, and sunspot rotations resulted in extreme flare activity (Wang et al., 2018).

A number of articles are devoted to analyses of the various properties and manifestations of the most significant flares. Huang, Xu, & Wang (2018) discussed the SOL2017-09-04 (M5.5) and SOL2017-09-06 (X9.3) flares from the perspective of the relationship between the intensity of the white-light emission and the proton flux of the SEPs near Earth. Romano et al. (2018) described two consecutive X-class flares that occurred on September 6 within a 3 hour time interval. Both of these flares displayed strong horizontal photospheric shearing motions and homologous white light ribbons (see also Verma, 2018; Yan et al., 2018a). Li et al. (2018), Seaton & Darnel (2018), Warren et al. (2018), and Yan et al. (2018b) studied the spectroscopic characteristics and evolution of the high-altitude coronal current sheet that formed in the wake of the halo CME associated with the near-the-west-limb SOL2017-09-10 (X8.2) flare.

As for space weather effects, general descriptions and solar sources are presented, for example, in Redmon et al. (2018) and Shen et al. (2018). The enhancements of the proton flux near Earth were considered in a number of publications. Using the SEPs that occurred on September 4, 6, and 10, Chertok (2018a,b) confirmed that the flux density and frequency spectrum of microwave radio bursts reflect the number and energy spectrum of accelerated particles, including the 10–100 MeV protons coming to Earth. Kurt et al. (2018) analyzed the 6–7% GLE connected with the SOL2017-09-10 (X8.2) flare and determined the proton release time and anisotropy of the GLE fluxes. Kataoka et al. (2018) estimated the maximum radiation dose rate during this GLE at an airliner flight altitude of 12 km. Schwadron et al. (2018) discussed these particle radiation enhancements from the perspective of future human deep-space missions. Luhmann et al. (2018) studied the roles of coronal pseudostreamers and interplanetary shock magnetic connectivity in the widespread multipoint SEP events of September 2017.

The present paper addresses other space weather disturbances caused by the arrival of interplanetary magnetic clouds or ICMEs to Earth. In Section 2, following the method of early diagnostics of solar eruption geoeffectiveness suggested by Chertok et al. (2013, 2015), Chertok, Grechnev, & Abunin (2017), we select the extreme ultraviolet (EUV) arcade and dimming regions of the SOL2017-09-04 (M5.5) and SOL2017-09-10 (X8.2) eruptive flares and calculate their magnetic fluxes. In Section 3, we use the calculated eruptive fluxes as initial parameters for posterior estimates of the expected amplitude of Forbush decreases and geomagnetic storms (Dst, *A*p indexes) and compare them with actual observations. We also consider the surrounding circumstances under which these eruptions occurred, in particular, the influence of a high-speed solar wind emanating from adjacent coronal hole CH823 on the propagation of the CMEs/ICMEs from these eruptions and the interaction of two ICMEs near Earth. Then, in Section 4, a discussion of the results and conclusions are presented.

## 2 Magnetic flux of arcades and dimmings in the EDSEG tool

Eruptions of a large CME are accompanied on the solar disk by large-scale phenomena such as post-eruption (PE) arcades and dimmings. Dimmings are regions where the EUV (and soft X-ray) brightness of coronal structures is temporarily reduced during an ejection and persist over several hours (Thompson et al., 1998; Hudson & Cliver, 2001; Dissauer et al. (2018) and references therein). The deepest stationary long-lived core dimmings adjacent to the eruption center are interpreted to be mainly the result of plasma outflow and density decreases in the footpoints of erupting and expanding CME flux ropes. Large-scale arcades of bright loops growing in size over time arise in the region of the main body of pre-eruption magnetic flux ropes ejected as a part of CMEs (Kahler, 1977; Sterling et al., 2000; Gopalswamy et al., 2017). As a whole, PE arcades and dimmings represent the structures and areas involved in the CME process.

Chertok et al. (2013, 2015) studied approximately 50 events in which major nonrecurrent geomagnetic storms (GMSs) and Forbush decreases (FDs) with solar eruptive sources in the central part of the disk are sufficiently reliably identified**.** They showed that the total (summarized) unsigned magnetic flux of the longitudinal field at the photospheric level within the arcade and dimming areas is a suitable quantitative parameter for the earliest diagnostics of the geoefficiency of a solar eruption, i.e., for estimates of the probable intensities (scales) of a GMS and an FD and their onset and peak times. The arcades and dimmings are formed at the initial stage of an eruption and visualize the structures involved in the CME process, while the measurements of the CME velocity and other its parameters are carried out later. Importantly, these earliest estimates can be carried out immediately after an eruption without information on a CME. We call this technique the early diagnostics of solar eruption geoeffectiveness (EDSEG tool).

This tool was developed by Chertok et al. (2013, 2015) based on the analysis of eruptions and associated geospace disturbances observed during 1996 – 2005 with the *Extreme-ultraviolet Imaging Telescope* (EIT) and the *Michelson Doppler Imager* (MDI) onboard the *Solar and Heliospheric Observatory* (SOHO; Domingo, Fleck, & Poland, 1995). After 2010 – 2011, matched data regarding solar eruptions have been obtained with the EUV telescope *Atmospheric Imaging Assembly* (AIA; Lemen et al., 2012) in the 193 Å channel and the magnetograph *Helioseismic and Magnetic Imager* (HMI; Scherrer et al., 2012), both onboard the *Solar Dynamic Observatory* (SDO; Pesnell, Thompson, & Chamberlin, 2012). The transition procedure from the SOHO to SDO data is elaborated in Chertok, Grechnev, & Abunin (2017) using the relatively short overlapping intervals for EIT–AIA and MDI–HMI detailed observations.  Additionally, a number of large eruptions over the next five years, with 12-hour cadence EIT images, were compared against AIA for the transition effort. The extracted arcade and dimming areas essentially coincide in the SOHO/EIT 195 Å and SDO/AIA 193 Å images. However, the SOHO/MDI line-of-sight magnetic flux systematically exceeds the SDO/HMI flux in the same areas by a factor of 1.4.

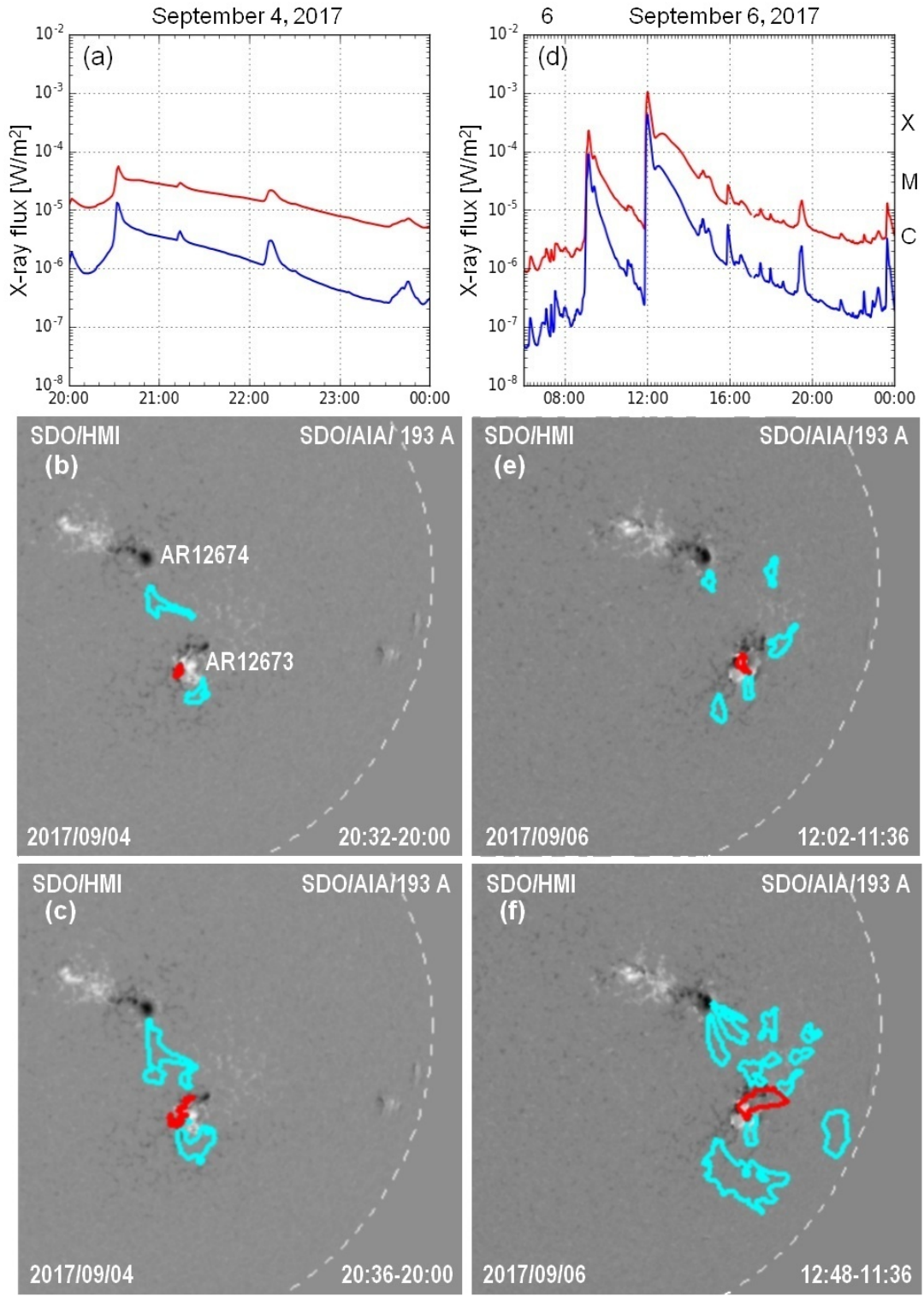

Figure 1. Data from the flares and eruptions on September 4 (left row) and 6 (right row), 2017. (a, d) GOES soft X-ray time profiles. (b, c, e, f) SDO/HMI magnetograms of the central-southwest sector of the disk with superposed dimming (blue) and arcade (red) contours extracted from the AIA 193 Å fixed-base difference images. Active regions AR12673 and AR12674 are marked in panel (b).

Naturally, in this study of the September 2017 events, we used the SDO data considering the transition procedure outlined above. The cross-calibration factor between the EUV telescopes SOHO/EIT and SDO/AIA, required for analysis of the September 2017 events, was close to 3.0.

For Figure 1 we extract the arcades and dimmings from the SDO/AIA images with the same techniques as those used in Chertok et al. (2013, 2015) and Chertok, Grechnev, & Abunin (2017), based on formal criteria referring mainly to the relative brightness variations. For arcades, a criterion that extracted the area around the flare site where the brightness exceeded 5% of the maximum during the event was appropriate. A brightness depression deeper than 40% of the pre-event level is an optimal criterion for the extraction of significant core dimmings located near the eruption center and obviously related to the eruption. The area of a post-eruption arcade increases over time. To avoid ambiguity, we extracted an arcade area from an image of each event that is temporally close to the maximum EUV flux from the selected area. Usually, this time is close to or somewhat later than the peak time of a corresponding GOES soft X-ray flare. The parameters of the dimmings were computed from the so-called "portrait", which shows all dimmings appearing during the analyzed time interval of the event in a single composite image. The "dimming portrait" is generated by determining the minimum brightness (the maximum dimming depth) in each pixel over the entire analyzed fixed-base difference set. It is these dimming "portraits" that are displayed with blue contours in Figure 1. The total unsigned eruptive magnetic flux within the extracted arcade and dimming areas is evaluated at the photospheric level by co-aligning the resulting AIA difference images with an HMI line-of-sight magnetogram recorded just before an eruption. The corresponding AIA and HMI FITS files were downloaded from the Stanford Joint Science Operations Center (http://jsoc2.stanford.edu/data/aia/synoptic/ and http://jsoc2.stanford.edu/data/hmi/fits/).

Among the events in early September 2017, two flares were geoeffective: SOL2017-09-04T20:33 (hereafter referred to as UT) (M5.5, coordinates S06W15) and SOL2017-09-06T12:02 (X9.3, S09W38). Their development can be seen in the AIA daily movies at https://sdo.gsfc.nasa.gov/data/dailymov.php. Only these two flares were accompanied by Earth-directed partial halo CMEs (https://cdaw.gsfc.nasa.gov/CME_list/HALO/halo.html) and large arcade and dimming eruptive magnetic fluxes (Figure 1). The third halo-associated flare SOL2017-09-10 was near-the-limb and could not have a strong impact on the Earth's environment except for an SEP/GLE effect. The remaining powerful flares corresponded with only insignificant CMEs and small arcade and dimming areas. These insignificant effects apply in particular, to the first of two consecutive X-class flares on September 6 SOL2017-09-06T09:10 (X2.2) and to the X-class flare on September 7 SOL2017-09-07T14:36 (X1.1).

In Figure 1, the upper row displays the GOES soft X-ray time profiles of the two geoeffective flares SOL2017-09-04T20:33 and SOL2017-09-06T12:02. The two lower rows show the central-SW sector of the disk with the arcades and dimmings extracted from the AIA 193 Å fixed-base difference images of these flares against the backgrounds of the SDO/HMI magnetograms. Here, the extracted arcades and dimmings are outlined by red and blue contours, respectively. We used the extraction procedure and criteria described above in this section. In both eruptive flares, the arcade areas were rather compact and were located directly over the sunspots of AR12673. At the same time, the configuration and extent of the dimmings show how large-scale structures were involved in both eruptions, especially the second one. At the moment of the main soft X-ray peak, rather large dimming patches stretched far from AR12673. The arcades and dimmings fully developed later when their extracted areas significantly increased. A comparison with pre-eruptive AIA images, presented in Figures 2a and 2b, shows that the northern dimmings correspond to transequatorial loops connecting AR12673 and AR12674. In addition, large loop structures located to the south and west of AR12673 were involved in the eruption process, especially associated with the SOL2017-09-06T12:02 flare (Figures 1f and 2b).

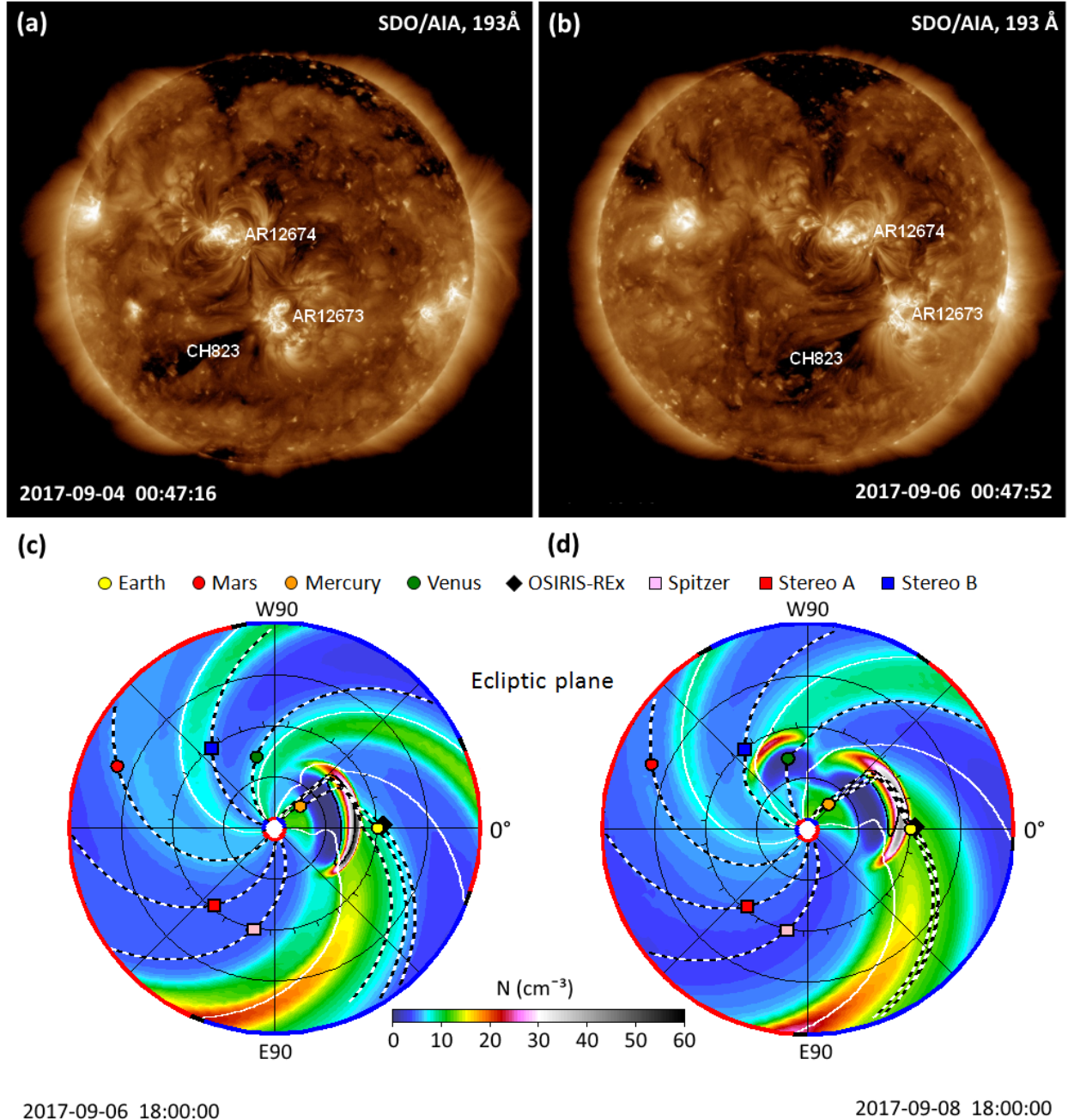

**Figure 2.** (a, b) Pre-eruptive SDO/AIA images from the 193 Å channel showing northern AR12674, main flare-productive AR12673 and adjacent extended CH823. (c, d) Frames of the WSA-ENLIL+Cone modeling with solar wind number density illustrating the propagation of the ICMEs from the flares on September 4 (left row) and 6 (right row), 2017 within a high-speed solar wind stream flowing from CH823.

As a result, at 20:36, the total eruptive magnetic flux at the photospheric level was $\Phi \approx 53.5$ mfu (here and elsewhere, magnetic flux is expressed in units of 1 mfu = $10^{20}$ Mx) in the SOL2017-09-04 flare (Figure 1c). For the SOL2017-09-06T12:02 (X9.3) flare, we calculated the eruptive flux at the peak of the delayed soft X-ray component, i.e., at 12:48 (see Figure 1d, f). At this time, the eruptive flux was $\Phi \approx 111.3$ mfu, i.e., twice as much as the SOL2017-09-04 flare. In both flares, the estimated arcade and dimming fluxes are approximately equal. These equal fluxes occur because the relatively small arcades are located in strong magnetic fields directly over AR12673, whereas the dimmings, although they occupy much larger areas, are associated with weaker field structures.

## 3 Space Weather Consequences

Once the total magnetic fluxes of the arcades and dimmings in the two eruptions are known, we can estimate the possible scales of the geospace effects caused by the arrival of the corresponding ICMEs to Earth with the EDSEG tool. This tool has at least four limitations. First, we consider the eruptions that generate isolated CMEs/ICMEs that propagate in the background solar wind without interacting with other transients. Second, the EDSEG tool works provided that the ejecta will hit Earth and, for GMSs, the $Bz$ component in the leading, sheath, body or trailing part of an ICME is predominantly negative (southern), which is not predicted by our tool. This outcome follows from Chertok et al. (2013); from the very beginning, we considered eruptions that were sources of GMSs that had already taken place and were therefore due to ICMEs with a negative $Bz$ component. This limitation means that the tool can give the earliest estimates of probable FD magnitude (which does not depend on the sign of Bz) and of the approximate maximum possible GMS scale. Third, we are referring to the earliest diagnostics of solar eruptions, not estimates of the exact parameters of GMSs and FDs but evaluations of the scales of these upcoming space weather disturbances. Fourth, for eruptions occurring in the central zone of the visible solar hemisphere within ±45°of the disk center, we did not consider projection effects. All these restrictions are justified for the earliest tentative estimates of the scales of geospace disturbances.

In Chertok et al. (2013, 2015), the EDSEG empirical relationships were determined from the SOHO data by connecting the total unsigned magnetic fluxes of the arcades and dimmings with the probable magnitudes of the GMSs and FDs and the time intervals between the peak of the associated soft X-ray burst with the start ($\Delta$To) and peak ($\Delta$Tp) of the corresponding GMS. Chertok, Grechnev, & Abunin (2017) adapted these relationships to the contemporary SDO data as described in Section 2. For the total magnetic fluxes of arcades and dimmings ($\Phi$) measured with the SDO/HMI magnetometer, the EDSEG empirical relationships are as follows ($\Phi$ is in mfu, i.e., $10^{20}$ Mx):

- For the GMS intensity (Dst and *A*p indexes)

$$\text{Dst [nT]} = 30 - 15.4(\Phi + 3.8)^{1/2},$$
$$Ap\ [2\text{nT}] = 1.12\ \Phi.$$

- For the FD magnitude

$$FD[\%] = -0.3 + 0.042\Phi.$$

- For the onset ($\Delta$To) and peak ($\Delta$Tp) transit times, i.e., the intervals between the eruption time (time of the associated soft X-ray burst maximum) and the onset and peak of the corresponding GMS

$$\Delta To\ [h] = 98/(1+0.00616\ \Phi),$$
$$\Delta Tp\ [h] = 118/(1+0.0056\ \Phi).$$

The results of applying these relationships to the two eruptive events under consideration are presented in Table 1 and will be discussed below in comparison with the actual geospace disturbances.

**Table 1**

*Estimated and Observed Parameters of Space Weather Disturbances from Two Eruptive Flares in September 2017*

| Parameters | SOL2017-09-04T20:33 | | SOL2017-09-06T12:02 | |
|---|---|---|---|---|
| | Estimated | Observed | Estimated | Observed |
| FD [%] | 1.9 | 2.0 | 4.4 | 9.3 |
| Dst [nT] | –87 | –23 | –135 | –144, –124 |
| $Ap$ [2 nT] | 60 | 18 | 125 | 205, 235 |
| $\Delta$To [h] | 74 | 50 | 58 | 37 |
| $\Delta$Tp [h] | 91 | 61.5 | 73 | 40, 53 |

Before this discussion, it is reasonable to consider the circumstances under which the two eruptions in question occurred. The SDO/AIA 193 Å images presented in Figures 2a and 2b demonstrated that the parent AR12673 was connected with the northern AR12674 by large-scale transequatorial loops.

In the immediate vicinity and somewhat to the west of AR12673, extended coronal hole CH823 was located. This coronal hole was in an Earth facing position on September 4–5. There are reasons to believe that the Earth-directed sectors of both halo CMEs from September 4 and 6 were propagated along the high-speed stream (HSS) emanating from CH823 first in the corona and then in interplanetary space, and their aerodynamic drags were consequently relatively small. According to the SOHO/LASCO movies (https://lasco-www.nrl.navy.mil/daily_mpg/), in the field of view of the coronagraph C3 at the positional angles PA ≈ 200–220°, both transients had sufficiently high plane-of-the-sky speeds; the September 4 CME had shock and ejecta plane-of-the-sky speeds of approximately 1650 and 1550 km/s, respectively, and the September 6 CME moved in the corona with somewhat lower speeds of 1500 and 1300km/s, respectively. The WSA-ENLIL+Cone modeling (see Wold et al., 2018) shown in Figures 2c and 2d (https://iswa.gsfc.nasa.gov/IswaSystemWebApp/) also indicates the possible propagation of both transients inside the HSS from CH823. Due to these factors, ICMEs can arrive at Earth faster than usual.

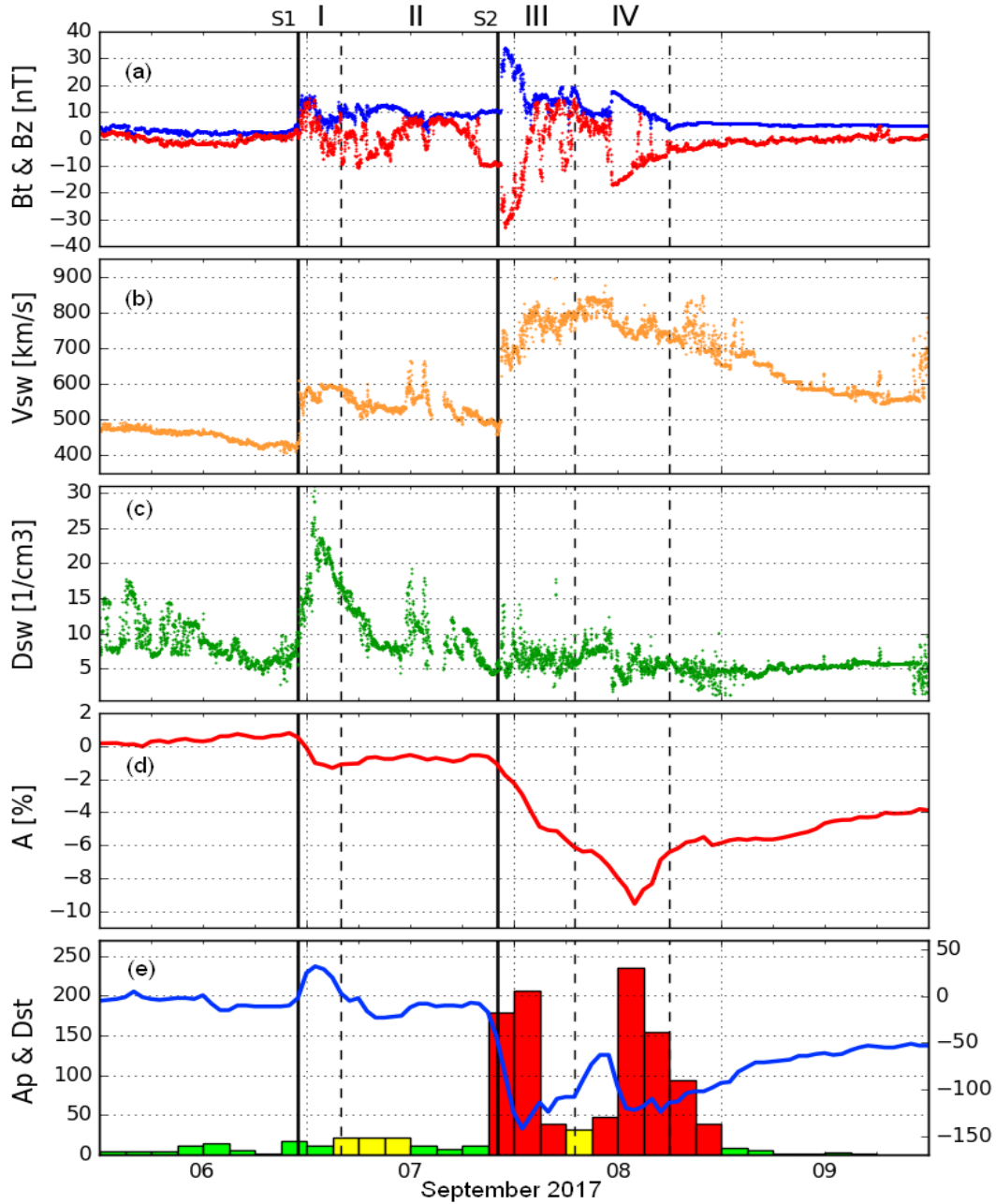

**Figure 3.** Geospace data for September 6–9,2017. (a–c) *In situ* DSCOVR measurements of the total magnetic field *Bt* (blue), *Bz* component (red), and solar wind speed and density near Earth. (d, e) Forbush decreases and geomagnetic disturbances (Dst and *Ap* indexes). A pair of solid lines correspond to the shocks driven by the first and the second ICMEs and feature abrupt and simultaneous increases in magnetic magnitude and solar wind speed; regions I and III bound the sheaths, and regions II and IV mark the ejecta.

The analysis of the geospace events and parameters of September 4–8 showed that the ICMEs under consideration interacted not only with a quiet solar wind but also with each other, with the HSS from CH823, with a corotating interaction region (CIR) on the western edge, and finally with a heliospheric current sheet (HCS), which on these days was not far from the eruption area, CH or Earth. These factors complicated the structures and affected the amplitudes of the space weather disturbances, as seen in Figure 3. The time profiles of the-near-Earth total strength of the interplanetary magnetic field (IMF), $Bz$ component, and solar wind speed and density are taken from the *Deep Space Climate Observatory* (DSCOVR; https://www.ngdc.noaa.gov/dscovr/portal/index.html#/vis/summary). The values of the hourly storm-time disturbance index Dst, calculated from the data of four low-latitude geomagnetic observatories and characterizing the effect of the global equatorial ring current, are presented at http://wdc.kugi.kyoto-u.ac.jp/dstdir/index.html. The linear planetary three-hour *A*p index is defined as the mean value of the variations of the terrestrial magnetic field, which corresponds to a Kp index, and is calculated using data from geomagnetic stations located at moderately high geomagnetic latitudes mainly in the northern hemisphere. The data are downloaded from ftp://ftp.gfz-potsdam.de/pub/home/obs/kp-ap/wdc/. The maximum FD magnitude is adopted from the IZMIRAN Database of the Forbush-Effects and Interplanetary Disturbances (FEID; http://spaceweather.izmiran.ru/eng/dbs.html). This maximum value corresponds to a cosmic ray rigidity of 10 GV and is determined by data from the world network of neutron monitors using the global survey method (see Belov et al., 2018).

Figures 3a–3c demonstrate that the first ICME-driven shock from the SOL2017-09-04T20:33 eruption (S1) arrived at Earth at approximately midnight on September 6, i.e., with start transit time $\Delta$To $\approx$ 50 h, and brought an IMF with a positive $Bz \approx$ 15 nT (hourly values). In the subsequent sheath (region I), the magnetic field strength and sign were variable with a $Bz$ amplitude of approximately 10 nT. The ICME body or ejecta (region II) presented certain signatures of a magnetic cloud, where the negative $Bz$ component first switches to positive and then back to negative at the 10 nT level again. The shock of the second ICME (S2), associated with the SOL2017-09-06T12:02 (X9.3) flare, came to Earth at an interval of less than a day relative to the first transient with a start transit time of $\Delta$To $\approx$ 34 h, i.e., 16 h less than that of the shock of the first ICME. This lower start transit time occurred despite that in the corona, the September 6 CME was slightly slower than the September 4 CME, as noted above. This may be a consequence of the second ICME propagating in the rarefied interplanetary plasma in the wake of the first ICME, which resulted in the second ICME overtaking the trailing part of the first ICME before arriving at Earth at 22 UT on September 7 and forming a combined complex disturbance (see Shen et al., 2018). The shock from the second ICME compressed and reinforced magnetic field of the first ICME, which resulted in very large strengths for both $Bt$ and $Bz$ of up to 27.3 nT and –24.2 nT, respectively, in the sheath leading part (region III). Another persistent and strong negative $Bz$ of up to –18 nT occurred in the trailing sector of the ICME ejecta (region IV), which was not affected by various interactions, came to Earth without much distortion and where the signatures of a magnetic cloud became especially pronounced.

We did not make an actual forecast regarding the geospace disturbances caused by the eruptions of early September 2017. However, we are interested in analyzing the results the EDSEG tool yields if the earliest estimates of the disturbances are carried out realistically using the available operational data from the EUV images and magnetograms only, i.e., without any information on the CMEs.

For the SOL2017-09-04T20:33 (M5.5) flare, which occurred slightly to the west of the central meridian at heliolongitude W15, the relatively small eruptive arcade and dimming flux of $\Phi \approx$ 53.5 mfu leads to a probable GMS intensity of Dst $\approx -87$ nT and $Ap \approx 60$ (in units of 2 nT). These results indicate that only a weak to moderate GMS is expected from this eruption. The observational data, presented in Figures 3d, 3e and Table 1, reveal that the ICME-driven shock caused only a mechanical compression of the magnetosphere (a positive surge of Dst). A small subsequent decline to Dst $\approx -23$ nT and a very small increase to $Ap \approx 18$ were observed in the middle of September 7. As Figure 3a shows, the main reason for the lack of a significant GMS is the positive (northern) sign of the $Bz$ component at the ICME front. This reasoning is supported by the estimated FD amplitude (1.9%), which does not depend on the $Bz$ sign and is close to the observed value (2.0%). For the time parameters, the onset and peak of the space weather effects (mainly FD) were observed almost 1–1.5 days earlier than the estimates from the magnetic flux. This discrepancy apparently occurred because, as noted above, the halo CME from September 4 at PA $\approx 200$–$220°$ had a sufficiently high plane-of-the-sky speed in the corona (1550–1650 km/s) much higher than average, and the Earth-directed sector of this ICME, starting directly from the corona, propagated inside a high-speed solar wind flowing from coronal hole CH823 and was not subjected to significant drag.

The SOL2017-09-06T12:02 (X9.3) flare occurred at heliolongitude W38, and therefore, the line-of-sight eruptive flux could be underestimated due to the projection effect reducing the arcade and dimming areas. Nevertheless, the flux was $\Phi \approx 111$ mfu, i.e., approximately 2 times greater than that of the SOL2017-09-04 eruption. This flux indicates that, under favorable conditions, the eruption on September 6 could cause quite intense space weather effects, with an estimated GMS intensity of Dst $\approx -135$ nT, $Ap \approx 125$, an FD amplitude $\approx 4.4\%$ and time onset and peak intervals of $\Delta$To $\approx 58$ h and $\Delta$Tp $\approx 73$ h. From Figure 3 and Table 1, in this case again the ICME front arrived at Earth much earlier (at the end of September 7) due to sufficiently high initial CME speed in the corona (1300–1500 km/s), weak drag in the high-speed plasma from CH823 and a rarefied wake from the preceding ICME from the September 4 eruption. When comparing the expected intensity of the space weather disturbances with observations, their scale was reasonably estimated as moderate to strong. However, the real amplitude of the FD and GMS (by the $Ap$ index) was somewhat larger and more complicated. The combination of the strong IMF and high-speed solar wind is the main factor determining the magnitude of the FD in the form of a rapid decline in the cosmic ray density behind the second shock and the deep minimum of the FD inside the ejecta in the middle of September 8 when the amplitude reached 9.3%. There were two well-defined active periods of geomagnetic activity on September 8. At the beginning of the disturbance just after the shock front, the largest $Bz$ negative was presented, and the indexes were Dst $\approx -124$ nT and $Ap \approx 205$. In the ejecta, where the signs of the magnetic cloud with the large southern $Bz$ became pronounced again, these indexes were Dst $\approx -109$ nT and $Ap \approx 235$.

Noticeable geomagnetic activity could have occurred in connection with the first ICME, but at the moment when $Bz$ became large and negative on September 7, the second ICME-driven shock arrived, and the GMS capability was instead generated by this second event. Thus, the interaction of the two ICMEs increased the efficiency of the second one but perhaps reduced the efficiency of the first one.

It is reasonable to compare the posterior EDSEG estimates with the forecasts of the WSA–ENLIL+Cone model (see Wold et al., 2018). The latter is based on a much larger and later set of input data, including CME parameters such as speed, direction of propagation, as well as the background solar wind plasma and magnetic field. The forecasts, released several hours after eruptions, when information about CMEs became available, and presented at https://iswa.gsfc.nasa.gov/IswaSystemWebApp/, show that the application of the ENLIL model to the September events yields approximately the same results as the EDSEG tool. The ENLIL model clearly overestimates the scale of the GMS from the first SOL2017-09-04T20:33 eruption (Kp~7) and generally does not deal with the FD amplitude. Meanwhile, the relatively accurate EDSEG forecast of FD~1.9–2.0% (see Table 1) allows us to argue that the very weak GMS in this case is caused by the positive Bz-component. For the GMS initiated by the second SOL2017-09-06T12:02 eruption, both the EDSEG and ENLIL tools estimate its scale relatively correctly as strong with Kp~7 and Dst~ -135 nT and Ap~125 [2 nT], respectively. The predicted Earth arrival times for the two ICMEs indicated by both tools were much later than the observed ones, although the differences the ENLIL forecasts and the observed times were noticeably lower (approximately 9 and 16 h) than those of the EDSEG estimates (24 and 21 h).

Overall, the data presented above show that application of the EDSEG tool to the eruptions in September 2017 in general gives positive results. Despite the limitations of the tool outlined at the beginning of this section and the rather complex nature and structures of the resulting interplanetary transients (see Redmon et al., 2018; and Shen et al., 2018), the earliest diagnostics of the two eruptions by the total magnetic fluxes of the arcades and dimmings estimate the GMS and FD scales as weak–moderate for the first event and moderate–strong for the second event, which are close to the observed scales. For the time parameters of the geospace disturbances, their actual onset occurred much earlier than the estimates of the EDSEG tool. We argue above that this time discrepancy is mainly due to the propagation of both CMEs/ICMEs through the high-speed plasma stream from CH823 and the second ICME through the wake of the first ICME.

## 4 Discussion and Conclusions

The outstanding AR12673 was superflaring active. However, only two flares, SOL2017-09-04T20:33 (M5.5) and SOL2017-09-06T12:02 (X9.3), were accompanied by Earth-directed halo CMEs and were geoeffective in the sense of excitation of FDs and GMSs. These two ICMEs interacted with each other near Earth, as well as with the HSS from CH823, CIR, and a heliospheric current sheet. These interactions resulted in the combination of a strong IMF and high-speed solar wind near Earth, which was not observed in the 24th solar cycle until September 2017. The September 8 GMS was strong by the Dst and $Ap$ indexes but severe by the $Kp$ = 8+ (ftp://ftp.gfz-potsdam.de/pub/home/obs/kp-ap/wdc/). Large FDs, such as those observed on September 8, occur no more than once a year on average, and in the current cycle, only two events have been larger than the September event (in March 2012 and in June 2015).

The extraction of associated EUV arcades and dimmings and the calculation of their total unsigned magnetic fluxes yielded relatively correct posterior estimates for the probable scale of the disturbances using the EDSEG tool. For the September 4 event, this accuracy is evidenced by the estimated and observed FD magnitudes almost coinciding, whereas the measured GMS amplitude was greatly weakened due to the northern $Bz$ component in the ICME front near Earth.

The structure of the ICME from the September 6 eruption was complicated by interaction with the trailing part of the previous ICME and the simultaneous arrival of the fast solar wind and the CIR from CH823 at Earth. These interactions seem to be the main reason for certain discrepancies between the estimates of the power of the geospace disturbances and their actual magnitudes. The first FD on the morning of September 7 was least affected by this interaction and developed almost without interference of its minimum, so we have almost a complete coincidence of the estimated and observed FD amplitude. Unlike the FD, the geomagnetic activity created by the first ICME did not develop normally when the tail section of the first ICME with a negative $Bz$ approached Earth. For the second event, the interaction of the two ICMEs, starting on the night of September 7, greatly increased both the IMF module and its negative $Bz$ component. As a result, the actual intensities of both the FD and GMS (by the $Ap$ index) were somewhat greater than the estimated magnitudes, although comparable in scale. The general correspondence between the estimated and observed scales of the space weather disturbances means that the eruptive magnetic flux of arcades and dimmings is an important factor that largely determines the geoeffectiveness of the related CMEs/ICMEs.

In principle, an alternative interpretation of the September 7–8 geospace disturbances is that they correspond to the HSS stream that originated from the nearby CH823. The shock wave S2 reached Earth at the end of September 7 (see Figure 3), when the HSS could also come to Earth. Of course, the HSS contributed to the formation of this and subsequent composite disturbances. However, the combination of the solar (time, location and features of flare, eruption and halo CME) and interplanetary data and the data regarding the ICME propagation are in good agreement and indicate that these disturbances were mainly caused by the eruption on September 6 and not an HSS (see Shen et al., 2018). The amplitude of shock S2 is too large to be related to an HSS. The significant fluctuations in the magnetic field and speed profiles in region III correspond well with the ICME sheath. The duration of the whole disturbance (shock S2 plus regions III and IV) and the time of the ejecta (region IV) are within the range typical for ICMEs.

The SOL2017-09-04T20:33 (M5.5) and SOL2017-09-06T12:02 (X9.3) flares occurred in more or less similar conditions and locations as a result of shearing motion, rotation and interaction of the newly emerging bipolar patches near the main pre-existing leading spot of positive polarity (Yang et al., 2017). Nevertheless, the ICME shock from the first flare brought to Earth a magnetic field with a positive $Bz$ component, and the ICME from the second flare had a predominately negative one near Earth. The exact causes of this difference are unclear. A number of factors could affect the orientation of the $Bz$ component. Unlike the September 4 flare, the more powerful and extended SOL2017-09-06T12:02 (X9.3) flare had a precursor in the form of the rather intense compact X2.2 flare SOL2017-09-06T09:10. The SOL2017-09-04-associated CME/ICMEs erupted and propagated in rather undisturbed conditions, in particular, inside the high-speed flow from CH823, while the transient from the SOL2017-09-06T12:02 flare moved through the relatively small CME from the SOL2017-09-06T09:10 flare and through the wake of the first CME/ICME and interacted with them. Further, erupting flux ropes and CMEs/ICMEs as a whole can undergo significant rotation about their rise direction as they ascend in the corona and propagate in interplanetary space (see Manshester et al. (2017) for a review). This rotation is especially characteristic of eruptions from highly sheared ARs and when vortex flows take place at AR footpoints, as in these cases (Yang et al., 2017). All these factors could lead to different orientations of $Bz$ near Earth.

Quite often, a CME is deflected from a radial trajectory by the open field of an adjacent CH (Gopalswamy, 2009; Manshester et al. 2017). In our case, CMEs from the two flares under consideration apparently were not strongly deflected by CH823. This lack of deflection especially applies to the powerful CME from September 6, which was associated with a flare at heliolongitude W38, and this CH was located on the path of this flare to Earth. Instead of deflection, another effect took place: the Earth-directed sectors of both CMEs appeared to be launched and propagated within the high-speed, low-density stream from CH823. Perhaps this effect explains why the observed transit times of both ICMEs were noticeably lower than our estimates and those from the WSA+ENLIL model (see Table 1 and Figure 2). The high initial velocities of the shocks and CMEs in the corona and the sequential propagation of the two ICMEs also contributed to the differences in transit times. In addition, the wind speed from CH823 was apparently also quite high; in the previous solar rotation it was approximately 600 km/s at 1 AU.

The relatively accurate diagnostic results for the September 2017 flares are further evidence that the EDSEG tool, despite a number of limitations outlined in the beginning of Section 3, is quite applicable for the earliest estimates of the probable scale of geospace disturbances driven by actual solar eruptions. Clearly, this tool should be used as a starting component with a combination of other methods for short-term GMS and FD forecasting, including those based on measurements of near-the-Sun CMEs, MHD models, and stereoscopic observations of ICME propagation (see Manchester et al., 2017; Webb & Nitta, 2017). Real-time forecasting could thus begin with the EDSEG tool from only near-solar-surface manifestations of an earthward eruption and then could be specified by other techniques as suitable additional data becomes available over the course of development.

In addition, the analysis of the eruptions and disturbances of September 2017 demonstrates the important roles of two effects. The interaction of ICMEs, especially occurring near Earth and coinciding with the passage of a high-speed stream, co-rotating interaction region or heliospheric current sheet, can strongly affect geospace parameters. The presence of an extended CH near the parent AR can lead not only to CME deflection but also to lesser ICMEs being dragged within an HSS and to an earlier onset of FDs and GMSs. For a more detailed understanding of these events, further studies are required.

### Acknowledgments

We are grateful to two anonymous reviewers and to the AGU Space Weather Editorial Office, whose suggestions have improved this article. The authors thank the SDO/AIA and HMI, NOAA/SWPC GOES, DSCOVR, SOHO/LASCO, WSA+ENLIL, and other related teams for the open data used in this study. All data sources used in producing the results presented in this article are quoted in sections 2 and 3. The AIA and HMI FITS files were downloaded from the Stanford Joint Science Operations Center (http://jsoc2.stanford.edu/data/aia/synoptic/ and http://jsoc2.stanford.edu/data/hmi/fits/). Data for the GOES soft X-ray time profiles were taken at ftp://ftp.swpc.noaa.gov/pub/lists/xray/. The in situ DSCOVR data were obtained from https://www.ngdc.noaa.gov/dscovr/portal/index.html#/vis/summary. Data for the Dst and Ap geomagnetic indexes are available on the websites http://wdc.kugi.kyoto-u.ac.jp/dstdir/index.html and ftp://ftp.gfz-potsdam.de/pub/home/obs/kp-ap/wdc/. The maximum Forbush decrease magnitude was adopted from http://spaceweather.izmiran.ru/eng/dbs.html. We are thankful to V.V. Grechnev for developing the procedures for extraction of the arcade and dimming areas and calculation of their magnetic flux. This research was partially supported by the Russian Foundation of Basic Research under grants 17-02-00308 and 17-02-00508 and by the Russian Science Foundation under grant 15-12-20001.

**References**

Attie, R., Kirk, M.S., Thompson, B.J., Muglach, K., & Norton, A.A. (2018). Precursors of magnetic flux emergence in the moat flows of active region AR12673. *Space Weather*, *16*(8), 1143–1155. https://doi.org/10.1029/2018SW001939

Belov, A., Eroshenko, E., Yanke, V., Oleneva, V., Abunin, A., Abunina, M., Papaioannou, A., & Mavromichalaki, E. (2018). The global survey method applied to ground level cosmic ray measurements. *Solar Physics, 293*(4), id. 68.https://doi.org/10.1007/s11207-018-1277-6

Chertok, I.M. (2018a). Powerful solar flares of 2017 September: correspondence between parameters of microwave bursts and proton fluxes near Earth. *Research Notes of the AAS* , *2*, id. 20. https://doi.org/10.3847/2515-5172/aaaab7

Chertok, I.M. (2018b). Diagnostic analysis of the solar proton flares of September 2017 by their radio bursts. *Geomagnetism and Aeronomy, 58*(4), 457–46. https://doi.org/10.1134/S0016793218040035

Chertok, I.M., Abunina, M.A., Abunin, A.A., Belov, A.V., & Grechnev, V.V. (2015). Relationship between the magnetic flux of solar eruptions and the Ap index of geomagnetic storms. *Solar Physics, 290*(2), 627–633. https://doi.org/10.1007/s11207-014-0618-3

Chertok, I.M., Grechnev, V.V., & Abunin, A.A. (2017). An early diagnostics of the geoeffectiveness of solar eruptions from photospheric magnetic flux observations: the transition from SOHO to SDO. *Solar Physics, 292*(4), id. 62. https://doi.org/10.1007/s11207-017-1081-8

Chertok, I.M., Grechnev, V.V., Belov, A.V., & Abunin, A.A. (2013). Magnetic flux of EUV arcade and dimming regions as a relevant parameter for early diagnostics of solar eruptions - sources of non-recurrent geomagnetic storms and Forbush decreases. *Solar Physics, 282*(1), 175–199. https://doi.org/10.1007/s11207-012-0127-1

Dissauer, K., Veronig, A.M., Temmer, M., Podladchikova, T., & Vanninathan, K. (2018). On the detection of coronal dimmings and the extraction of their characteristic properties. *Astrophysical Journal, 855(2), id. 137.*https://doi.org/10.3847/1538-4357/aaadb5

Domingo, V., Fleck, B., & Poland, A.I. (1995). The SOHO mission: an overview. *Solar Physics, 162*(1–2), 1–37. https://doi.org/10.1007/BF00733425

Gopalswamy, N., Makela, V., Xie, H., Akiyama, S., & Yashiro, S. (2009). CME interactions with coronal holes and their interplanetary consequences. *Journal of Geophysical Research,114*(3), id. A22. https://doi.org/10.1029/2008JA013686

Gopalswamy, N., Yashiro, S., Akiyama, S., & Xie, H. (2017). Estimation of reconnection flux using post-eruption arcades and its relevance to 1-AU magnetic clouds. *Solar Physics, 292*(4), id. 65. https://doi.org/10.1007/s11207-017-1080-9

Huang, N., Xu, Y., & Wang, H. (2018). Relationship between intensity of white-light flares and proton flux of solar energetic particles. *Research Notes of the AAS* , *2*, id. 7. https://doi.org/10.3847/2515-5172/aaa602

Hudson, H.S. & Cliver, E.W. (2001). Observing coronal mass ejections without coronagraphs. *Journal of Geophysical Research*, *106*(A11), 25199–25214. https://doi.org/10.1029/2000JA904026

Kahler, S (1977). The morphological and statistical properties of solar X-ray events with long decay times. *Astrophysical Journal, 214*, 891–897. https://doi.org/10.1086/155319

Kataoka, R., Sato, T., Miyake, S., Shiota, D., & Yûki Kubo, Y. (2018). Radiation Dose Nowcast for the Ground Level Enhancement on 10-11 September 2017. *Space Weather*, *16*(7), 917–923.https://doi.org/10.1029/2018SW001874

Kurt, V., Belov, A., Kudela, K., & Yushkov, B. (2018). Some characteristics of GLE on 2017 September 10. *Contributions of the Astronomical Observatory Skalnate Pleso*, *48*, 329–338. (DOI will will appear in the near future.)

Lemen, J.R., Title, A.M., Akin, D.J., Boerner, P.F., Chou, C., Drake, J.F., … Waltham, N. (2012). The Atmospheric Imaging Assembly (AIA) on the Solar Dynamics Observatory (SDO), *Solar Physics,275*(1–2), 17–40. https://doi.org/10.1007/s11207-011-9776-8

Li, Y., Xue, J.C., Ding, M.D., Cheng, X., Su, Y., Feng, L., … Gan, W.Q. (2018). Spectroscopic observations of a current sheet in a solar flare. *Astrophysical Journal Letters*, *853*(1), id. L15. https://doi.org/10.3847/2041-8213/aaa6c0

Luhmann, J.G., Mays, M.L., Li, Y., Lee, C.O., Bain, H., Odstrcil, D., Mewaldt, R.A., Cohen, C.M.S., Larson, D., &Petrie, G. (2018). Shock Connectivity and the Late Cycle 24 Solar Energetic Particle Events in July and September 2017. *Space Weather, 16*(5), 557–568. https://doi.org/10.1029/2018SW001860

Manchester IV, W., Kilpua, E. K. J., Liu, Y. D. , Lugaz, N., Riley, P., Török, T., &Vršnak, B. (2017). The physical processes of CME/ICME evolution. *Space Science Reviews*, *212*(3–4), 11591159–1219. https://doi.org/10.1007/s11214-017-0394-0

Pesnell, W. D., Thompson, B.J., & Chamberlin, P.C. (2012). The Solar Dynamics Observatory (SDO). *Solar Physics,275*(1–2), 3–15. https://doi.org/10.1007/s11207-011-9841-3

Redmon, R.J., Seaton, D.B., Steenburgh, R., He, J., & Rodriguez, J.V. (2018). September 2017's Geoeffective Space Weather and Impacts to Caribbean Radio Communications during Hurricane Response. *Space Weather,*       https://doi.org/10.1029/2018SW001897

Romano, P., Elmhamdi, A., Falco, M., Costa, P., Kordi, A.S., Al-Trabulsy, H.A., & Al-Shammari, R.M. (2018). Homologous white light solar flares driven by photospheric shear motions. *Astrophysical Journal Letters*, *852*(1), id. L10. https://doi.org/10.3847/2041-8213/aaa1df

Seaton, D.B., & Darnel, J.M. (2018). Observations of an eruptive solar flare in the extended EUV solar corona. *Astrophysical Journal Letters*, *852*(1), id. L9. https://doi.org/10.3847/2041-8213/aaa28e

Scherrer, P.H., Schou, J., Bush, R.I., Kosovichev, A. G., Bogart, R. S., Hoeksema, J. T, … Tomczyk, S. (2012). The Helioseismic and Magnetic Imager (HMI) investigation for the Solar Dynamics Observatory (SDO). *Solar Physics, 275*(1–2), 207–227. https://doi.org/10.1007/s11207-011-9834-2

Schwadron, N.A., Rahmanifard, F., Wilson, J., Jordan, A.P., Spence, H.E., Joyce, C. J., … Zeitlin, C. (2018). Update on the worsening particle radiation environment observed by CRaTER and implications for future human deep-space exploration. *Space Weather*, *16*(3), 289–304.https://doi.org/10.1002/2017SW001803

Shen, C., Xu, M., Wang Y., Chi, Y., & Luo, B. (2018). Why the Shock-ICME Complex Structure is Important: Learning From the Early 2017 September CMEs. *Astrophysical Journal, 861*(1), id. 28. https://doi.org/10.3847/1538-4357/aac204

Sterling, A.C., Hudson, H.S., Thompson, B.J., & Zarro, D. (2000). Yohkoh SXT and SOHO EIT Observations of Sigmoid-to-Arcade Evolution of Structures Associated with Halo Coronal Mass Ejections.*Astrophysical Journal,532*(1), 628–647. https://doi.org/10.1086/308554

Sun, X., & Norton, A.A. (2017). Super-flaring active region 12673 has One of the fastest magnetic flux emergence ever observed. *Research Notes of the AAS, 1*, id. 24. https://doi.org/10.3847/2515-5172/aa9be9

Thompson, B.J., Plunkett, S.P., Gurman, J.B., Newmark, J.S., St. Cyr, O.C., & Michels, D.J. (1998). SOHO/EIT observations of an Earth-directed coronal mass ejection on May 12, 1997. *Geophysical Research Letters,25*(14), 2465–2468. https://doi.org/10.1029/98GL50429

Verma, M. (2018). On the origin of two X-class flares in active region NOAA 12673 – Shear flows and head-on collision of new and pre-existing flux. *Astronomy & Astrophysics*, *612*, id. A101.  https://doi.org/10.1051/0004-6361/201732214

Wang, H., Yurchyshyn, V., Liu, C., Ahn, K., Toriumi, S., & Cao, W. (2018). Strong transverse photosphere magnetic fields and twist in light bridge dividing delta sunspot of active region 12673. *Research Notes of the AAS,2*, id. 8. https://doi.org/10.3847/2515-5172/aaa670

Warren, H.P., Brooks, D.H., Ugarte-Urra, I., Reep, J.W., Crump, N.A., & Doschek, G.A. (2018). Spectroscopic observations of current sheet formation and evolution. *Astrophysical Journal,854*(2), id.122. https://doi.org/10.3847/1538-4357/aaa9b8

Webb, D. & Nitta, N. (2017). Understanding problem forecasts of ISEST campaign flare-CME events. *Solar Physics, 292*(10), id. 142. https://doi.org/10.1007/s11207-017-1166-4

Wold, A.M., Mays, M.L., Taktakishvili, A., Jian, L.K., Odstrcil, D., &MacNeice, P. (2018). Verification of real-time WSA-ENLIL+Cone simulations of CME arrival-time at the CCMC from 2010 to 2016. *Journal of Space Weather and Space Climate, 8*, id. A17. https://doi.org/10.1051/swsc/2018005

Yan, X.L., Wang, J.C., Pan, G.M., Kong, D.F., Xue, Z.K., Yang, L.H., Li, Q.L., & Feng, X.S. (2018a). Successive X-class flares and coronal mass ejections driven by shearing motion and sunspot rotation in active region NOAA 12673. *Astrophysical Journal, 856*(1), id. 79. https://doi.org/10.3847/1538-4357/aab153

Yan, X.L., Yang, L.H., Xue, Z.K., Mei, Z.X., Kong, D.F., Wang, J.C., & Li, Q.L. (2018b). Simultaneous observation of a flux rope eruption and magnetic reconnection during an X-class solar flare. *Astrophysical Journal Letters*, *853*(1), id. L18. https://doi.org/10.3847/2041-8213/aaa6c2

Yang, S., Zhang, J., Zhu, X., & Song, Q. (2017). Block-induced complex structures building the flare-productive solar active region 12673.*Astrophysical Journal Letters*, *849*(2), id. L21. https://doi.org/10.3847/2041-8213/aa9476